\documentclass[aps,pre,twocolumn,floatfix]{revtex4}

\usepackage{graphicx}
\usepackage{amssymb,amsfonts,amsmath}
\usepackage{epsf}
\usepackage{subfigure}
\usepackage{epstopdf}

\newcommand{\be}{\begin{equation}}
\newcommand{\ee}{\end{equation}}
\newcommand{\bea}{\begin{eqnarray}}
\newcommand{\eea}{\end{eqnarray}}

\begin{document}

\title{\bf Kinetics and thermodynamics of DNA polymerases with exonuclease proofreading}

\author{Pierre Gaspard}
\affiliation{Center for Nonlinear Phenomena and Complex Systems,\\
Universit\'e Libre de Bruxelles, Code Postal 231, Campus Plaine,
B-1050 Brussels, Belgium}

\begin{abstract}
Kinetic theory and thermodynamics are applied to DNA polymerases with exonuclease activity, taking into account the dependence of the rates on the previously incorportated nucleotide.  The replication fidelity is shown to increase significantly thanks to this dependence at the basis of the mechanism of exonuclease proofreading.  In particular, this dependence can provide up to a hundred-fold lowering of the error probability under physiological conditions.  Theory is compared with numerical simulations for the DNA polymerases of T7 viruses and human mitochondria.
\end{abstract}

\noindent 
\vskip 0.5 cm

\maketitle

\section{Introduction}

In the companion paper~\cite{paperI}, the kinetic theory and thermodynamics of exonuclease-deficient (exo$^-$) DNA polymerases have been developed on the basis of experimental observations from biochemistry \cite{PWJ91,WPJ91,DPJ91,J93,TJ06,JJ01a,JJ01b,LNKC01,LJ06,LK82,EG91,KB00,SUKOOBWFWMG08,RBT08,ZBNS09,DJPW10,BBT12} and analytical methods to solve the kinetic equations of copolymerization \cite{AG08,AG09,GA14}.  In this way, the error probability has been studied numerically and analytically for the exo$^-$ DNA polymerases of T7 viruses and human mitochondria, showing how replication fidelity is determined by kinetics and related to thermodynamics.

Molecular and thermal fluctuations at the nanoscale induce errors in DNA replication, at the origin of possible mutations.  Following the discovery and systematic experimental studies of DNA polymerases \cite{LBSK58,BLSK58,BK72}, Hopfield, Ninio, and Bennett have shown in the seventies how kinetics can reduce the error probability if replication is driven out of equilibrium \cite{H74,N75,B79}.  Thanks to the kinetic amplification of discrimination between correct and incorrect base pairs, exo$^-$ DNA polymerases can already lower their error probability down to values of about $10^{-5}$-$10^{-6}$.  However, the theory of quasispecies by Eigen and Schuster \cite{ES77,ES78a,ES78b} implies that the self-replication of a quasispecies requires that the mutation probability should be lower than a threshold inversely proportional to its genome size.  For genome sizes as large as $10^{10}$~nucleotides in higher eukaryotes \cite{GEFS09}, the mutation probability should thus be as low as $10^{-10}$.  Therefore, biological evolution towards such organisms would not have been possible without dedicated proofreading mechanisms greatly enhancing the fidelity of DNA replication.  Progress in molecular biology has revealed that proofreading is specifically generated, on the one hand, by the exonuclease activity of DNA polymerases, able to cleave incorrectly incorporated nucleotides one at a time, as well as by postreplication mismatch repair \cite{J93}.  This latter mechanism is the feature of other enzymes than DNA polymerases that will not be considered here \cite{L74,SR83,IPBM06}.

In the present paper, our goal is to extend the analysis of the companion paper~\cite{paperI} to DNA polymerases with exonuclease proofreading and to investigate the implications of the dependence of the rates on the previously incorporated nucleotide.  In the presence of the exonuclease activity, we shall see that this dependence becomes essential to significantly lower the error probability and increase fidelity in the transmission of genetic information.

In Section~\ref{Kin+Thermo}, the kinetic scheme is extended to include the reactions of the exonuclease activity.  The kinetic equations are explicitly given in Appendix~\ref{AppA}. As for exo$^-$ polymerases, these equations are reduced for the Michaelis-Menten kinetics, which continues to hold in the presence of the exonuclease activity.  The thermodynamics of the enzymatic activities is also presented.  The kinetic equations are solved analytically and the thermodynamic quantities are deduced under the assumptions of the Bernoulli-chain model in Section~\ref{B-chain} and Appendix~\ref{AppB}, and of the Markov-chain model in Section~\ref{M-chain} and Appendix~\ref{AppC}.  In Sections~\ref{T7-Pol} and~\ref{Hum-Pol}, the enzymatic process is numerically simulated for the exo$^+$ DNA polymerases of T7 viruses and human mitochondria and the results are analyzed with the theoretical methods.  Conclusions are drawn in Section~\ref{Conclusion}.

\section{Kinetics and thermodynamics}
\label{Kin+Thermo}

\subsection{Generalities}

Most DNA polymerases have an exonuclease proofreading mechanism besides their polymerase activity.  The polymerase and exonuclease activities may be on the same polypeptide of the protein complex forming the enzyme, or on separate subunits.  The elongation of DNA is catalyzed by the polymerase domain.  If this latter is slowed down by the insertion of an incorrect nucleotide, the growing copy moves to the exonuclease domain where the incorrect nucleotide is cleaved by hydrolysis.  The reactions are thus
\bea
&&\mbox{polymerase activity:}\nonumber\\ 
&&{\rm dNTP} \ + \ {\rm E}\cdot{\rm DNA}_l \  \quad \rightleftharpoons \quad
{\rm E}\cdot{\rm DNA}_{l+1} \ + \ {\rm PP}_{\rm i} \, ,
\label{pol-act}  \\
&&\mbox{exonuclease activity:}\nonumber\\ 
&&{\rm E}\cdot{\rm DNA}_{l+1} \ + \ {\rm H}_2{\rm O}  \quad \rightleftharpoons \quad
{\rm E}\cdot{\rm DNA}_l \ + \ {\rm dNMP} \, , \quad
\label{exo-act} \\
&&\mbox{overall reaction:}\nonumber\\ 
&& {\rm dNTP} \ + \ {\rm H}_2{\rm O}
  \qquad\quad \ \rightleftharpoons \quad
{\rm dNMP} \ + \ {\rm PP}_{\rm i} \, , 
\label{hydrolysis}
\eea
where $E$ denotes the enzyme, DNA$_l$ the deoxyribonucleic double helix with the template strand and the growing copy, dNTP deoxyribonucleoside triphosphate, PP$_{\rm i}$ pyrophosphate, and dNMP deoxyribonucleoside monophosphate.  The exonuclease activity proceeds by hydrolysis of the ultimate nucleotide attached to the growing chain, so that a deoxyribonucleoside monophosphate dNMP returns to the surrounding solution.  The overall reaction~(\ref{hydrolysis}) is the hydrolysis of dNTP into dNMP with the release of pyrophosphate by the polymerase activity.  The Guldberg-Waage condition for the chemical equilibrium of this overall reaction is given by
\be
\frac{[{\rm dNTP}]_{\rm eq} \, c^0}{[{\rm dNMP}]_{\rm eq} \, [{\rm PP}_{\rm i}]_{\rm eq}} = \exp\left(\frac{\Delta G^0}{RT}\right) 
\label{GW_eq}
\ee
in terms of the standard free enthalpy of hydrolysis into pyrophosphate $\Delta G^0 = -45.6\, {\rm kJ/mol}$ \cite{CQ09,MRMGPGB03,FA95}, the standard concentration $c^0=1$~M, the molar gas constant $R=8.31451$~J~K$^{-1}$~mol$^{-1}$, and the temperature $T$.

Given the physiological concentrations \cite{T94,H01}:
\bea
&&[{\rm dNTP}] \simeq \mbox{5-40 $\mu$M} \, , \label{physio-dNTP}\\
&&[{\rm dNMP}] \simeq \mbox{0.3-20 $\mu$M} \, , \label{physio-dNMP}\\
&&[{\rm PP}_{\rm i}] \simeq \mbox{0.2-0.3 mM} \, , \label{physio-PP}
\eea
the enzyme remains very much away from chemical equilibrium even if the mean growth velocity is vanishing.  Indeed, Eq.~(\ref{GW_eq}) implies exceedingly small dNTP concentrations for physiological dNMP concentrations, or unphysical dNMP concentrations far above one molar for physiological dNTP concentrations.  Contrary to exonuclease-deficient DNA polymerases where the vanishing of the growth velocity corresponds to thermodynamic equilibrium, this latter is not accessible if exonuclease is active, so that energy dissipation is always present and the enzyme remains out of equilibrium, unless it dissociates from DNA.

As in the companion paper~\cite{paperI}, the dissociation of the enzyme-DNA complex is neglected, which is justified if the processivity is high enough.  Moreover, the surrounding solution is assumed to be sufficiently large in order to keep constant the concentrations of the different substances during the growth of the copy.

In the regime of steady growth, the mean growth velocity $v$ is given by
\be
v=r^{\rm p}-r^{\rm x} \, ,
\label{v-p-x}
\ee
where $r^{\rm p}=\langle\dot N_{\rm PP_i}\rangle$ denotes the polymerase rate, i.e., the rate of pyrophosphate release, while $r^{\rm x}=\langle \dot N_{\rm dNMP}\rangle$ is the exonuclease rate, i.e., the rate of dNMP release.  Consequently, the exonuclease rate is equal to the polymerase rate $r^{\rm p}=r^{\rm x}$ if the growth velocity is vanishing $v=0$.  Besides, the release of dNTP in the surrounding solution has the rate
\be
\langle \dot N_{\rm dNTP}\rangle = -r^{\rm p}-r^{\rm x} \, .
\label{r_dNTP}
\ee
Consequently, dNTP continues to be consumed if the growth velocity is zero, unless both the polymerase and exonuclease rates vanish, which only happens at the inaccessible chemical equilibrium.  We should thus expect that the thermodynamic entropy production remains positive in the presence of the exonuclease activity.

\subsection{Kinetic scheme}

\begin{figure}[h]
\centerline{\scalebox{0.65}{\includegraphics{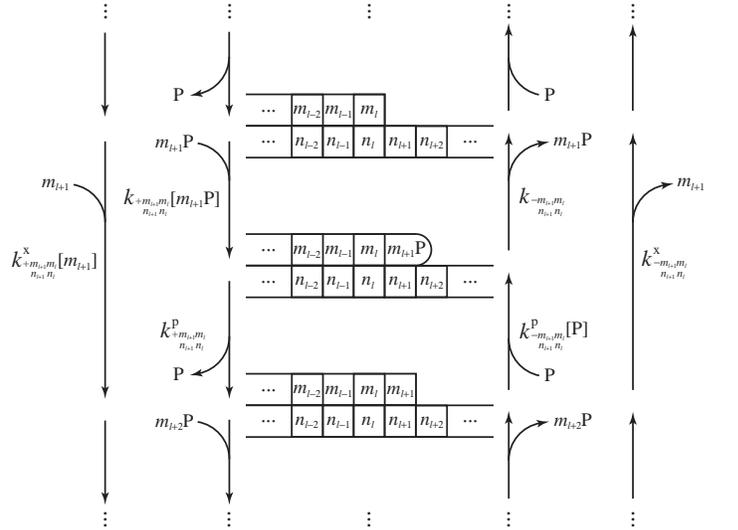}}}
\caption{Kinetic scheme of the polymerase and exonuclease activities.  $\{m_j\}$ denotes the ssDNA copy, $\{n_j\}$ the ssDNA template, $m_j{\rm P}$ deoxynucleoside triphosphates dNTP, $m_j$ deoxynucleoside monophosphates dNMP, and P pyrophosphates PP$_{\rm i}$.}
\label{fig1}
\end{figure}

Here, we consider the kinetic scheme depicted in Fig.~\ref{fig1}.  The reaction rates of the elementary steps are determined by the mass action law, giving the appropriate dependence of the rates on the concentrations of nucleotides and pyrophosphate.  The template is represented by its sequence $\alpha=n_1\cdots n_l n_{l+1}\cdots$ and the copy by $\omega=m_1\cdots m_l$, where the successive nucleotides are denoted $m,n\in\{ {\rm A}, {\rm C}, {\rm G}, {\rm T}\}$.  In this notation, we can express the dependence of the kinetics on the local environment of the ultimate nucleotide being attached or detached.

The reactions of the polymerase activity are the same as in the companion paper~\cite{paperI}.  From a copy ending with the ultimate monomeric unit $m_l$, the binding and dissociation of the deoxyribonucleoside triphosphate $m_{l+1}{\rm P}$ have the rates
\bea
\mbox{dNTP binding:} \qquad &&k_{+m_{l+1} m_l\atop \ \, n_{l+1} \, n_l}[m_{l+1}{\rm P}] \, ,
\label{ntb_rate} \\
\mbox{dNTP dissociation:} \qquad &&k_{-m_{l+1} m_l\atop \ \, n_{l+1} \, n_l} \, .
\label{ntd_rate}
\eea
From the copy ending with $m_{l+1}{\rm P}$, the release of pyrophosphate PP$_{\rm i}$ -- denoted P -- and the reverse reaction of pyrophosphorolysis have the rates
\bea
\mbox{polymerization:} \qquad &&k^{\rm p}_{+m_{l+1} m_l\atop \ \, n_{l+1} \, n_l} \, ,
\label{pol_rate} \\
\mbox{depolymerization:} \qquad &&k^{\rm p}_{-m_{l+1} m_l\atop \ \, n_{l+1} \, n_l}[{\rm P}] \, .
\label{depol_rate}
\eea
In the presence of the exonuclease activity, there are two further reactions.  In Fig.~\ref{fig1}, the binding of the deoxyribonucleoside monophosphate $m_{l+1}$ is depicted on the left-hand side, and its dissociation by hydrolysis on the right-hand side.  These reactions have the rates:
\bea
\mbox{dNMP binding:} \qquad &&k^{\rm x}_{+m_{l+1} m_l\atop \ \, n_{l+1} \, n_l}[m_{l+1}] \, ,
\label{nmb_rate} \\
\mbox{dNMP dissociation:} \qquad &&k^{\rm x}_{-m_{l+1} m_l\atop \ \, n_{l+1} \, n_l} \, ,
\label{nmd_rate}
\eea
this latter defining the rate constant of the exonuclease activity.

The kinetic equations of this scheme are given as Eqs.~(\ref{kin_eq_1})-(\ref{kin_eq_2}) in Appendix~\ref{AppA}. 

The template sequence is characterized by the probability distributions $\nu_l(\alpha)=\nu_l(n_1\cdots n_l)$ to find a subsequence $n_1\cdots n_l$ of length $l$.  The known properties of such sequences have been discussed in the companion paper~\cite{paperI}.  For reasons of simplicity, we suppose in the following that the template sequence is Bernoullian with $\nu_l(\alpha)=1/4^l$ for $l=1,2,3,...$.

\subsection{Michaelis-Menten kinetics}
\label{MM_kin}

Experimental observations \cite{PWJ91,WPJ91,DPJ91,J93,TJ06,JJ01a,JJ01b,LNKC01,LJ06,LK82,EG91,KB00,SUKOOBWFWMG08,RBT08,ZBNS09,DJPW10,BBT12} show that the binding and dissociation of deoxyribonucleoside triphosphates is faster than the other reactions:
\bea
&& k_{+m m' \atop \ \, n \, n'}[m{\rm P}], \ k_{-m m'\atop \ \, n \, n'} \gg \nonumber\\ 
&& k^{\rm p}_{+m m' \atop \ \, n \, n'}, \ k^{\rm p}_{-m m'\atop \ \, n \, n'}[{\rm P}], \ k^{\rm x}_{+m m' \atop \ \, n \, n'}[m], \ k^{\rm x}_{-m m'\atop \ \, n \, n'} \, .
\label{MM-hyp}
\eea
Accordingly, the sequences $m_1\cdots m_{l-1}m_l$ and $m_1\cdots m_{l-1}m_lm_{l+1}{\rm P}$ remain in quasi-equilibrium and the kinetic equations are simplified as shown in Appendix~\ref{AppA} into an equation for the time evolution of the probability:
\be
P_t(\omega\vert\alpha) = P_t\left(m_1 \cdots m_l \qquad\quad\ \atop n_1\, \cdots \, n_l \, n_{l+1}\cdots\right) \, ,
\label{prob}
\ee
which is the sum of probabilities~(\ref{prob_sum}), as in the companion paper~\cite{paperI}.  Now, the transition rates are given by the sums of the rates of polymerase and exonuclease activities:
\bea
W_{+m_{l+1} m_l\atop \ \, n_{l+1}\, n_l} &=& W^{\rm p}_{+m_{l+1} m_l\atop \ \, n_{l+1}\, n_l}+W^{\rm x}_{+m_{l+1} m_l\atop \ \, n_{l+1}\, n_l} \, , \label{W+} \\
W_{\quad\, -m_l m_{l-1}\atop n_{l+1}\, n_l \, n_{l-1}} &=& W^{\rm p}_{\quad\, -m_l m_{l-1}\atop n_{l+1}\, n_l \, n_{l-1}} + W^{\rm x}_{\quad\, -m_l m_{l-1}\atop n_{l+1}\, n_l \, n_{l-1}} \, . \label{W-}
\eea
The rates of polymerase activity have already been presented in the companion paper~\cite{paperI}, while those of exonuclease activity are the following ones.  The rate of dNMP binding is equal to
\be
W^{\rm x}_{+m_{l+1} m_l\atop \ \, n_{l+1}\, n_l}\equiv 
\frac{k^{\rm x}_{+m_{l+1} m_l\atop \ \, n_{l+1} \, n_l}[m_{l+1}]}
{Q_{\ \ \ \ \ m_l\atop n_{l+1} \, n_l}}
\label{Wx+}
\ee
and the rate of dNMP dissociation by the exonuclease is
\be
W^{\rm x}_{\quad\, -m_l m_{l-1}\atop n_{l+1}\, n_l \, n_{l-1}}\equiv 
\frac{k^{\rm x}_{-m_l m_{l-1}\atop \ \, n_l \, n_{l-1}}}
{Q_{\ \ \ \ \ m_l\atop n_{l+1} \, n_l}}
\label{Wx-}
\ee
with the same denominators
\be
Q_{\ \ \ \ \ m_l\atop n_{l+1} \, n_l}\equiv 1 + \sum_{m_{l+1}} \frac{[m_{l+1}{\rm P}]}{K_{m_{l+1} m_l\atop n_{l+1} \, n_l}}
\label{denom}
\ee
and Michaelis-Menten dissociation constants
\be
K_{m m'\atop n \, n'}\equiv 
\frac{k_{-m m' \atop \ \, n \, n'}}
{k_{+m m'\atop \ \, n \, n'}} \, ,
\label{MM_csts}
\ee
as for the rates of polymerase activity~\cite{paperI}.
The rates~(\ref{Wx+}) and~(\ref{Wx-}) are written for the reactive events occurring to the sequence $m_1\cdots m_l$ of the copy growing on the sequence $n_1\cdots n_l n_{l+1}\cdots$ of the template.  We notice that the detachment of $m_l$ has a rate that depends not only on the template nucleotides $n_{l-1}$ and $n_l$ forming the base pairs $m_{l-1}$:$n_{l-1}$ and $m_l$:$n_l$, but also on the next template nucleotide $n_{l+1}$ because of the Michaelis-Menten kinetics.  The stochastic process ruled by the rates of polymerase and exonuclease activities can be simulated with Gillespie's algorithm \cite{G76,G77}.

Following a cycle that is closing after an overall reaction~(\ref{hydrolysis}), we obtain the following Guldberg-Waage conditions for chemical equilibrium:
\be
\frac{[m{\rm P}]_{\rm eq}\, c^0}{[m]_{\rm eq} \, [{\rm P}]_{\rm eq}} = 
c^0 \, \frac{k_{-m m'\atop \ \, n \, n'}\, k^{\rm p}_{-m m'\atop \ \, n \, n'}\, k^{\rm x}_{+m m'\atop \ \, n \, n'}}{k_{+m m'\atop \ \, n \, n'}\, k^{\rm p}_{+m m'\atop \ \, n \, n'}\, k^{\rm x}_{-m m'\atop \ \, n \, n'}}
= \exp\left(\frac{\Delta G^0}{RT}\right) .
\label{GW_eq_bis}
\ee
Combining this thermodynamic constraint with the same assumption of proportionality between the rates of polymerization and depolymerization as in the companion paper~\cite{paperI}
\be
K_{\rm P} \equiv \frac{k^{\rm p}_{+m m'\atop \ \, n \, n'}}{k^{\rm p}_{-m m'\atop \ \, n \, n'}} ,
\label{K_P-dfn}
\ee
we obtain the rate constants of dNMP binding~(\ref{nmb_rate}) given by
\be
k^{\rm x}_{+m m'\atop \ \, n \, n'}=k^{\rm x}_{-m m'\atop \ \, n \, n'} \, \frac{K_{\rm P}}{K_{m m'\atop n \, n'}\, c^0} \, \exp\left(\frac{\Delta G^0}{RT}\right) 
\label{k^x_+}
\ee
in terms of the exonuclease rate constants~(\ref{nmd_rate}), which have been measured experimentally \cite{DPJ91,J93,JJ01b}.

We shall also assume for simplicity that the nucleotide concentrations are all equal:
\bea
&&[{\rm dNTP}] \equiv [{\rm dATP}] = [{\rm dCTP}] = [{\rm dGTP}] = [{\rm dTTP}] \, , 
\label{dNTP}\\
&&[{\rm dNMP}] \equiv [{\rm dAMP}] = [{\rm dCMP}] = [{\rm dGMP}] = [{\rm dTMP}] \, . 
\qquad
\label{dNMP}
\eea

After a long enough time, the copolymerization process reaches a regime of steady growth since there is no termination.  In this regime, the mean growth velocity~(\ref{v-p-x}) becomes constant and the sequence of the growing copy takes stationary statistical properties described by the probability distribution $\mu_l(\omega\vert\alpha)$ to find the sequence $\omega=m_1 \cdots m_l$ of length~$l$ given that the template has the sequence $\alpha$.  This distribution describes in particular the mismatches between the copy and the template, which are generated with the error probability $\eta$ \cite{paperI}.

\subsection{Thermodynamics and sequence disorder}

In the regime of steady growth, the entropy production -- which is explicitly given by Eq.~(\ref{entr-A}) in Appendix~\ref{AppA} -- can be written as \cite{AG08}
\be
\Sigma \equiv \frac{1}{R}\frac{d_{\rm i}S}{dt}= v \, A \geq 0  \qquad\mbox{with} \qquad A = \epsilon + D(\omega\vert\alpha) \, ,
\label{entr-prod}
\ee
in terms of the mean growth velocity $v$, the entropy production per nucleotide or affinity~$A$, the free-energy driving force per nucleotide~$\epsilon$, and the conditional Shannon disorder~$D(\omega\vert\alpha)$ per nucleotide in the sequence \cite{paperI}.  If replication fidelity is high enough and the substitutions are equiprobable, the conditional disorder per nucleotide can be estimated as
\be
D(\omega\vert\alpha) \simeq \eta \, \ln \frac{3{\rm e}}{\eta} \ll 1
\label{D-estim}
\ee
in terms of the error probability $\eta \ll 1$.

\section{Bernoulli-chain model}
\label{B-chain}

\subsection{Kinetics and error probability}

The simplest model assumes that the rates only depend on the nucleotide that is attached or detached and on whether the pairing is correct or incorrect.  Besides the polymerization and depolymerization rates $W^{\rm p}_{\pm{\rm c}}$ and $W^{\rm p}_{\pm{\rm i}}$ already presented in the companion paper~\cite{paperI}, the rates of the exonuclease activity (and its reverse) are given by
\bea
&&W^{\rm x}_{+{\rm c}} = \frac{k^{\rm x}_{+{\rm c}} \, [{\rm dNMP}]}{Q} \, , \quad
W^{\rm x}_{+{\rm i}} = \frac{k^{\rm x}_{+{\rm i}} \, [{\rm dNMP}]}{Q} \, , \label{Wx+_B}\\
&&W^{\rm x}_{-{\rm c}} = \frac{k^{\rm x}_{-{\rm c}}}{Q} \, , \ \ \qquad\qquad
W^{\rm x}_{-{\rm i}} = \frac{k^{\rm x}_{-{\rm i}}}{Q}\, , \label{Wx-_B}
\eea
with the Michaelis-Menten denominator:
\be
Q = 1 + \left(\frac{1}{K_{\rm c}}+\frac{3}{K_{\rm i}}\right) [{\rm dNTP}] \, ,
\label{MM-denom_B}
\ee
and the rate constants of dNMP binding
\bea
&& k^{\rm x}_{+{\rm c}} = k^{\rm x}_{-{\rm c}} \, \frac{K_{\rm P}}{K_{\rm c}\, c^0} \, \exp\left(\frac{\Delta G^0}{RT}\right) \, , \\
&& k^{\rm x}_{+{\rm i}} = k^{\rm x}_{-{\rm i}} \, \frac{K_{\rm P}}{K_{\rm i}\, c^0} \, \exp\left(\frac{\Delta G^0}{RT}\right) \, ,
\eea
introduced with Eq.~(\ref{k^x_+}).

Under these assumptions, the probability distribution of the copy sequence factorizes into the probabilities to have correct or incorrect base pairs, which read
\be
\mu({\rm c})=1-\eta  \qquad\mbox{and} \qquad \mu({\rm i})=\eta/3
\label{B-prob}
\ee
in terms of the error probability $\eta$.  Consequently, the growing copy is a Bernoulli chain.  The probabilities~(\ref{B-prob}) are here given by
\bea
\mu({\rm c}) &=& \frac{W^{\rm p}_{+{\rm c}}+W^{\rm x}_{+{\rm c}}}{W^{\rm p}_{-{\rm c}}+W^{\rm x}_{-{\rm c}}+v}  \, , \label{B_mu_c}\\
\mu({\rm i}) &=& \frac{W^{\rm p}_{+{\rm i}}+W^{\rm x}_{+{\rm i}}}{W^{\rm p}_{-{\rm i}}+W^{\rm x}_{-{\rm i}}+v}  \, , \label{B_mu_i}
\eea
where $v$ is the mean growth velocity.  This latter can be expressed as
\be
v = \frac{W^{\rm p}_{+{\rm c}}+W^{\rm x}_{+{\rm c}}}{1-\eta} - W^{\rm p}_{-{\rm c}} - W^{\rm x}_{-{\rm c}} = 3\, \frac{W^{\rm p}_{+{\rm i}}+W^{\rm x}_{+{\rm i}}}{\eta} - W^{\rm p}_{-{\rm i}} - W^{\rm x}_{-{\rm i}}
\label{B-velocity}
\ee
by using Eq.~(\ref{B-prob}) \cite{AG09}.  The error probability $\eta$ can thus be obtained as a root of a polynomial of degree two.  Besides, the polymerase and exonuclease rates are given by
\be
r^{\rho} = \nu^{\rho} \left[ W^{\rho}_{+{\rm c}}-W^{\rho}_{-{\rm c}}(1-\eta) + 3\, W^{\rho}_{+{\rm i}}- W^{\rho}_{-{\rm i}}\, \eta \right]
\label{rates-p-x_B}
\ee
with the stoichiometric coefficients $\nu^{\rm p}=+1$ and $\nu^{\rm x}=-1$ respectively for $\rho=$~p and~x.  We notice that the mean growth velocity~(\ref{v-p-x}) is recovered by Eqs.~(\ref{B-velocity}).

\subsection{Thermodynamics and sequence disorder}

As shown in Appendix~\ref{AppB}, the thermodynamic entropy production is indeed given by Eq.~(\ref{entr-prod}) in terms of the mean growth velocity $v$, the free-energy driving force $\epsilon$, and the conditional Shannon disorder per nucleotide $D(\omega\vert\alpha)$.  This latter takes the same expression in terms of the error probability as for the Bernoulli-chain model of exo$^-$ DNA polymerases \cite{paperI}, which is approximated by Eq.~(\ref{D-estim}) if $\eta\ll 1$.

\subsection{Low speed regime}

As aforementioned, the polymerase and exonuclease activities do not approach thermodynamic equilibrium if the mean growth velocity is vanishing $v=0$.  Instead, the polymerase and exonuclease rates become equal by Eq.~(\ref{v-p-x}) and they can be evaluated as
\be
r^{\rm p}_0 = r^{\rm x}_0 \simeq k^{\rm x}_{-{\rm c}} \, .
\label{rates_p+x_0_B}
\ee
Indeed, the exonuclease rate is given by Eq.~(\ref{rates-p-x_B}) with $\rho=$~x and
only the term with $W^{\rm x}_{-{\rm c}}$ dominates, since the rates of dNMP attachment $W^{\rm x}_{+{\rm c}}$ and $W^{\rm x}_{+{\rm i}}$ are very small by the Guldberg-Waage condition~(\ref{GW_eq}) while the term with $W^{\rm x}_{-{\rm i}}\eta$ is negligible because the error probability is also very small $\eta\ll 1$.  Moreover, the denominator~(\ref{MM-denom_B}) becomes unity in this regime where the dNTP concentration is small with respect to the Michaelis-Menten dissociation constants $K_{\rm c}$ and $K_{\rm i}$.

Now, setting the velocity equal to zero in Eqs.~(\ref{B-velocity}) and evaluating the different terms, we similarly obtain the critical value of dNTP concentration and the corresponding error probability as
\bea
&&[{\rm dNTP}]_{0,{\rm B}} \simeq K_{\rm c} \left( \frac{[{\rm P}]}{K_{\rm P}} + \frac{k^{\rm x}_{-{\rm c}}}{k^{\rm p}_{+{\rm c}}}\right) \, , \label{dNTP-0-B}\\
&& \eta_{0,{\rm B}} =3 \, \frac{k^{\rm p}_{+{\rm i}} \, [{\rm dNTP}]_{0,{\rm B}}}{k^{\rm x}_{-{\rm i}}\, K_{\rm i}} \simeq 3 \, \frac{k^{\rm p}_{+{\rm i}} \, K_{\rm c}}{k^{\rm x}_{-{\rm i}}\, K_{\rm i}} \left( \frac{[{\rm P}]}{K_{\rm P}} + \frac{k^{\rm x}_{-{\rm c}}}{k^{\rm p}_{+{\rm c}}}\right) , \qquad 
\label{eta-0-B}
\eea
 in the Bernoulli-chain model.  If $k^{\rm x}_{-{\rm c}}=0$, we recover an estimation of the equilibrium dNTP concentration given in Ref.~\cite{paperI} for this model.

The entropy production can also be evaluated from Eq.~(\ref{entr-B}) when the growth velocity is zero to get
\be
\frac{1}{R}\frac{d_{\rm i}S}{dt}\Big\vert_{0,{\rm B}} \simeq  k^{\rm x}_{-{\rm c}} \, \ln \frac{K_{\rm c}\, c^0 \, {\rm e}^{-\beta\Delta G^0}}{K_{\rm P}[{\rm dNMP}]}\, ,
\label{entr-0-B}
\ee
with $\beta=(RT)^{-1}$.  Since the entropy production is not vanishing, the polymerase remains out of equilibrium due to the exonuclease activity.  We notice that the entropy production~(\ref{entr-0-B}) would be infinite if the dNMP concentration was zero because the reverse exonuclease reaction would have zero probability to occur in such a fully irreversible regime.

\begin{figure}[h]
\centerline{\scalebox{0.7}{\includegraphics{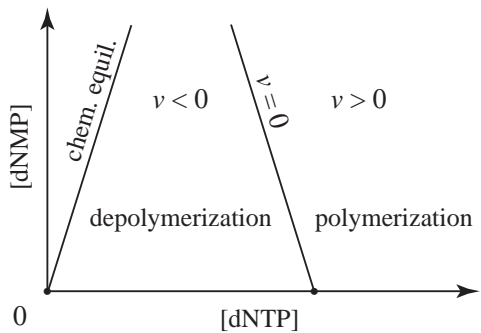}}}
\caption{Schematic diagram of enzymatic regimes in the plane of dNTP and dNMP concentrations showing the transition between polymerization and depolymerization if the mean growth velocity $v$ is vanishing, and the line where the reaction dNTP+H$_2$O~$\rightleftharpoons$~dNMP+PP$_{\rm i}$ is at chemical equilibrium.}
\label{fig2}
\end{figure}

In Fig.~\ref{fig2}, the behavior of DNA polymerases is schematically depicted in the plane of dNTP and dNMP concentrations.  In this plane, the chemical equilibrium condition~(\ref{GW_eq}) is a straight line going up from the origin with a very high slope.  Without any approximation, the condition of zero velocity obtained from Eqs.~(\ref{B-velocity}) would read $\gamma\, [{\rm dNTP}]_0 + \chi\, [{\rm dNMP}]=1$ with two positive coefficients $\gamma$ and $\chi$, which corresponds to the line $v=0$ in Fig.~\ref{fig2}.  Typically, the coefficients are ordered as $\gamma\gg\chi$ so that the approximation $[{\rm dNTP}]_0\simeq \gamma^{-1}$ given by Eq.~(\ref{dNTP-0-B}) is well satisfied.  The DNA copy is growing by polymerization at higher dNTP concentrations $[{\rm dNTP}]>[{\rm dNTP}]_0$, while it undergoes depolymerization for lower values $[{\rm dNTP}]<[{\rm dNTP}]_0$.

The complete thermodynamic equilibrium would be reached if both rates~(\ref{v-p-x}) and~(\ref{r_dNTP}) were vanishing, in which case $r_{\rm eq}^{\rm p}=r_{\rm eq}^{\rm x}=0$.  This would happen at the intersection of both oblique lines in Fig.~\ref{fig2}, but this point is at too large dNMP concentration to be accessible, confirming that the exonuclease activity keeps the enzyme away from equilibrium.

\subsection{Full speed regime}

The growth velocity tends to its maximal value if the dNTP concentration is larger than the Michaelis-Menten crossover concentration:
\be
[{\rm dNTP}] \gg \left(\frac{1}{K_{\rm c}}+\frac{3}{K_{\rm i}}\right)^{-1} \, .
\ee
In this regime, the detachment rates are negligible in Eqs.~(\ref{B-velocity}) so that the mean growth velocity and the error probability are given by
\bea
&& v_{\infty,{\rm B}} \simeq \frac{k^{\rm p}_{+{\rm c}}}{\displaystyle 1+3\, \frac{K_{\rm c}}{K_{\rm i}}}\, ,\label{v-infty-B}
\\
&& \eta_{\infty,{\rm B}} \simeq 3 \, \frac{k^{\rm p}_{+{\rm i}} \, K_{\rm c}}{k^{\rm p}_{+{\rm c}} \, K_{\rm i}} \, . \label{eta-infty-B}
\eea
If we compare with the results for exo$^-$ polymerases \cite{paperI}, we notice that these quantities are not affected by the exonuclease activity in the full speed regime.  The reason is that the exonuclease rate rapidly decreases as
\be
r^{\rm x}_{\rm B} \simeq \frac{k^{\rm x}_{-{\rm c}}}{[{\rm dNTP}]} \left(\frac{1}{K_{\rm c}}+\frac{3}{K_{\rm i}}\right)^{-1}\, , \label{r^x-infty-B}
\ee
if the dNTP concentration increases, so that the polymerase activity dominates: $v_{\infty}\simeq r_{\infty}^{\rm p}$.  Accordingly, the entropy production, the affinity, and the free-energy driving force increase in this regime as the logarithm of the dNTP concentration, in the same way as for exo$^-$ polymerases \cite{paperI}.

\section{Markov-chain model}
\label{M-chain}

\subsection{Kinetics and error probability}

Experimental observations show that the rates of DNA polymerases depend on both the newly and previously incorporated nucleotides \cite{J93}.  In the companion paper~\cite{paperI}, we have already given the polymerization-depolymerization rates.  With the exonuclease activity, we need to include also the corresponding rates
\bea
&& W^{\rm x}_{+p\vert p'} = \frac{k^{\rm x}_{+ p\vert p'} \, [{\rm dNMP}]}{Q_{p'}} \, , \label{Wx+_M} \\
&& W^{\rm x}_{-p\vert p'} = \frac{k^{\rm x}_{- p\vert p'}}{Q_{p}} \, , \label{Wx-_M}
\eea
with $p,p'=$~c or~i, whether the pair is correct or incorrect.
Here, we shall assume for simplicity that the exonuclease rates do not depend on the previously incorporated nucleotide, i.e., $k^{\rm x}_{- p\vert p'}=k^{\rm x}_{- p}$ for all $p'$.  However, the dependence on $p$ and $p'$ remains for the polymerase rates so that the Guldberg-Waage chemical equilibrium conditions~(\ref{GW_eq_bis}) give the rate constants
\be
k^{\rm x}_{+p\vert p'}=k^{\rm x}_{-p} \, \frac{K_{\rm P}}{K_{p\vert p'}\, c^0} \, \exp\left(\frac{\Delta G^0}{RT}\right) 
\label{k^x_M}
\ee
for $p,p'=$~c or~i.  These rates are thus determined in terms of available experimental data \cite{DPJ91,J93,JJ01b}.

These assumptions imply that a copy growing on a Bernoullian template is a Markov chain with conditional, tip, and bulk probabilities calculated as explained in the companion paper~\cite{paperI} in terms of the transition rates~(\ref{W+})-(\ref{W-}).  First, we need to calculate the partial velocities by iterating the self-consistent equations
\bea
v_{\rm c} &=& a\, v_{\rm c} + b\, v_{\rm i} \, , \label{v_c-M}\\
v_{\rm i} &=& c\,  v_{\rm c} + d\, v_{\rm i} \, , \label{v_i-M}
\eea
with the coefficients 
\bea
a &=& \frac{W_{+{\rm c}\vert{\rm c}}}{W_{-{\rm c}\vert{\rm c}} + v_{\rm c}} \, , \label{a-M} \\
b &=& \frac{3\, W_{+{\rm i}\vert{\rm c}}}{W_{-{\rm i}\vert{\rm c}} + v_{\rm i}} \, , \label{b-M}\\
c &=& \frac{W_{+{\rm c}\vert{\rm i}}}{W_{-{\rm c}\vert{\rm i}} + v_{\rm c}} \, , \label{c-M}\\
d &=& \frac{3\, W_{+{\rm i}\vert{\rm i}}}{W_{-{\rm i}\vert{\rm i}} + v_{\rm i}} \, , \label{d-M}
\eea
depending themselves on the partial velocities.  Thereafter, the tip probabilities are obtained by solving the following set of linear equations:
\bea
\mu({\rm c}) &=& a\, \mu({\rm c}) + 3\, c \, \mu({\rm i}) \, , \label{mu_c-M}\\
\mu({\rm i}) &=& \frac{b}{3}\, \mu({\rm c}) + d\, \mu({\rm i}) \, , \label{mu_i-M}
\eea
with the same coefficients~(\ref{a-M})-(\ref{d-M}), which are now determined.
The mean growth velocity is then given by
\be
v=v_{\rm c}\mu({\rm c})+3\, v_{\rm i}\mu({\rm i}) \, .
\label{M-velocity}
\ee
The conditional probabilities of the Markov chain can also be obtained~\cite{paperI}.

In the Markov-chain model, the polymerase and exonuclease rates can be expressed in terms of the conditional probabilities $\mu(p'\vert p)$ and the tip probabilities $\mu(p)$ of the Markov chain as
\be
r^{\rho} = \nu^{\rho}\, \sum_{p,p'}\left[W^{\rho}_{+p\vert p'} \, \mu(p') - W^{\rho}_{-p\vert p'}\, \mu(p'\vert p)\, \mu(p) \right] 
\label{rates_p+x_M}
\ee
with $\rho=$~p or x, the sum extending to $p,p'\in\{{\rm c},{\rm i},{\rm i},{\rm i}\}$, and the same stoichiometric coefficient $\nu^{\rho}$ as in Eq.~(\ref{rates-p-x_B}).

Now, the error probability is defined as $\eta\equiv 3 \, \bar\mu({\rm i})$ in terms of the bulk probability of incorrect base pairs.  Since the bulk and tip probabilities are proportional to each other according to $\bar\mu({\rm i}) = \mu({\rm i})\, v_{\rm i}/v$, we find that the error probability reads
\be
\eta = \frac{1}{\displaystyle 1+\frac{v_{\rm c}\,\mu({\rm c})}{3\, v_{\rm i}\,\mu({\rm i})}} \, .
\label{eta_M}
\ee

\subsection{Thermodynamics and sequence disorder}

It is in Appendix~\ref{AppC} that the expression is given for the thermodynamic entropy production of the Markov-chain model in the regime of steady growth.  Again, this expression can be written in the form~(\ref{entr-prod}) in terms of the mean growth velocity~$v$, the free-energy driving force~$\epsilon$, and the conditional disorder per nucleotide given by
\be
D(\omega\vert\alpha) = -\sum_{p,p'} \mu(p'\vert p) \, \bar\mu(p) \, \ln \mu(p'\vert p) \geq 0 \, ,
\ee
where $\mu(p'\vert p)$ and $\bar\mu(p)$ are respectively the conditional and bulk probabilities of the Markov chain and the sum extends to $p,p'\in\{{\rm c},{\rm i},{\rm i},{\rm i}\}$.

\subsection{Low speed regime}

The critical dNTP concentration where the growth velocity is vanishing can be obtained by requiring that Eqs.~(\ref{mu_c-M})-(\ref{mu_i-M}) admit a non-zero solution.  The condition for this result can be expressed as $(a-1)(d-1)=bc$ in terms of the coefficients~(\ref{a-M})-(\ref{d-M}) with $v_{\rm c}=v_{\rm i}=0$ so that the mean growth velocity~(\ref{M-velocity}) is indeed zero.  We obtain the critical value:
\be
[{\rm dNTP}]_{0,{\rm M}} \simeq K_{{\rm c}\vert{\rm c}} \left( \frac{[{\rm P}]}{K_{\rm P}} + \frac{k^{\rm x}_{-{\rm c}}}{k^{\rm p}_{+{\rm c}\vert{\rm c}}}\right) \, , \label{dNTP-0-M}
\ee
which is similar to the value~(\ref{dNTP-0-B}) obtained for the Bernoulli-chain model.  At this critical concentration, the polymerase and exonuclease rates are again given by Eq.~(\ref{rates_p+x_0_B}) as in the Bernoulli-chain model, while the entropy production~(\ref{entr-M}) takes the value
\be
\frac{1}{R}\frac{d_{\rm i}S}{dt}\Big\vert_{0,{\rm M}} \simeq  k^{\rm x}_{-{\rm c}} \, \ln \frac{K_{{\rm c}\vert{\rm c}}\, c^0 \, {\rm e}^{-\beta\Delta G^0}}{K_{\rm P}[{\rm dNMP}]}\, ,
\label{entr-0-M}
\ee
with $\beta=(RT)^{-1}$ in the Markov-chain model.  This confirms that the polymerase remains away from equilibrium even if the growth velocity is zero.

If the dNTP concentration is larger than the critical value~(\ref{dNTP-0-M}) but lower than the Michaelis-Menten dissociation constants for $p=$~c and~i
\be
[{\rm dNTP}]_{0,{\rm M}} \ll [{\rm dNTP}] \ll \left(\frac{1}{K_{{\rm c}\vert p}}+\frac{3}{K_{{\rm i}\vert p}}\right)^{-1} ,
\label{dNTP-range}
\ee
the mean growth velocity is no longer vanishing and it is important to determine how it increases with the dNTP concentration.  Since the error probability is expected to be very small $\eta\ll 1$, the probability for a correct base at the tip of the growing copy is much larger than for an incorrect base pair, $\mu({\rm c})\simeq 1\gg \mu({\rm i})$.  In order to satisfy Eq.~(\ref{mu_c-M}), the coefficient $a$ should be very close to unity: $a\simeq 1$.  Consequently, Eq.~(\ref{a-M}) implies that the corresponding partial velocity is approximated by $v_{\rm c}\simeq W_{+{\rm c}\vert{\rm c}}-W_{-{\rm c}\vert{\rm c}}$.  This expression is typically dominated by the polymerization rate $W^{\rm p}_{+{\rm c}\vert{\rm c}}$.  At dNTP concentrations lower than the Michaelis-Menten dissociation constants, the denominator is close to the unit value $Q_{\rm c}\simeq 1$, whereupon the mean growth velocity~(\ref{M-velocity}) can be evaluated as
\be
v\simeq v_{\rm c} \simeq \frac{k^{\rm p}_{+{\rm c}\vert{\rm c}}}{K_{{\rm c}\vert{\rm c}}} \, [{\rm dNTP}] \, ,
\label{M-v-low}
\ee
in the range~(\ref{dNTP-range}).

Now, we turn to the error probability at low but non-vanishing growth velocity.  Introducing the ratios of tip probabilities and partial velocities
\be
x\equiv \frac{\mu({\rm c})}{3\, \mu({\rm i})} \qquad\mbox{and}\qquad y\equiv \frac{v_{\rm c}}{v_{\rm i}} \, ,
\ee
the error probability~(\ref{eta_M}) can be rewritten as $\eta=(1+xy)^{-1}$.  Taking the ratios of Eqs.~(\ref{v_c-M})-(\ref{v_i-M}) and~(\ref{mu_c-M})-(\ref{mu_i-M}), we obtain quadratic equations for $x$ and $y$ in terms of the coefficients~(\ref{a-M})-(\ref{d-M}):
\be
b\, x = c\, y = \frac{1}{2}\left[ a-d+ \sqrt{(a-d)^2+4bc} \right] \, .
\label{eq-x-y}
\ee
At dNTP concentrations lower than the Michaelis-Menten dissociation constants where $Q_{\rm c}\simeq Q_{\rm i}\simeq 1$, the coefficients can be evaluated as
\bea
&& b \simeq 3 \, \frac{k^{\rm p}_{+{\rm i}\vert{\rm c}}\, [{\rm dNTP}]}{k^{\rm x}_{-{\rm i}}\, K_{{\rm i}\vert{\rm c}}} \, , \label{b-estim}\\
&& c \simeq \frac{k^{\rm p}_{+{\rm c}\vert{\rm i}}\, K_{{\rm c}\vert{\rm c}}}{k^{\rm p}_{+{\rm c}\vert{\rm c}}\, K_{{\rm c}\vert{\rm i}}} \, , \label{c-estim}
\eea
and $d\ll a$, so that $a \simeq 1 \gg b,c,d$.  Replacing in Eq.~(\ref{eq-x-y}), we find that the error probability is approximated by $\eta \simeq (xy)^{-1} \simeq bc$, hence
\be
\eta_{\rm M} \simeq 3 \, \frac{k^{\rm p}_{+{\rm i}\vert{\rm c}}\, k^{\rm p}_{+{\rm c}\vert{\rm i}}\, K_{{\rm c}\vert{\rm c}}}{k^{\rm x}_{-{\rm i}}\, k^{\rm p}_{+{\rm c}\vert{\rm c}}\, K_{{\rm c}\vert{\rm i}}\, K_{{\rm i}\vert{\rm c}}}  \, [{\rm dNTP}] \, ,
\label{eta-low-M}
\ee
at low speed.

\subsection{Full speed regime}

At concentrations satisfying the conditions
\be
[{\rm dNTP}] \gg \left(\frac{1}{K_{{\rm c}\vert p}}+\frac{3}{K_{{\rm i}\vert p}}\right)^{-1}\quad\mbox{for}\ \ p={\rm c} \ \mbox{and}\ {\rm i} \, ,
\ee
the exonuclease activity characterized by the rate~(\ref{rates_p+x_M}) with $\rho=$~x decreases as
\be
r^{\rm x}_{\rm M} \simeq \frac{K_{{\rm c}\vert{\rm c}}}{[{\rm dNTP}]} \left( k^{\rm x}_{-{\rm c}}+ 3 \, k^{\rm x}_{-{\rm i}} \, \frac{k^{\rm p}_{+{\rm i}\vert{\rm c}}\, K_{{\rm c}\vert{\rm i}}}{k^{\rm p}_{+{\rm c}\vert{\rm i}}\, K_{{\rm i}\vert{\rm c}}}\right)
\label{r^x-infty-M}
\ee
for increasing dNTP concentration.  Accordingly, the polymerase activity dominates in this regime and we recover the same expressions of the mean growth velocity, the error probability, the entropy production, the affinity, and the free-energy driving force per nucleotide, as for exonuclease-deficient polymerases~\cite{paperI}.  In particular, the error probability is given by
\be
\eta_{\infty,{\rm M}} \simeq 3 \, \frac{k^{\rm p}_{+{\rm i}\vert{\rm c}} K_{{\rm c}\vert{\rm c}}}{k^{\rm p}_{+{\rm c}\vert{\rm c}}K_{{\rm i}\vert{\rm c}}} \, ,
\label{eta-infty-M}
\ee
which is comparable to the result~(\ref{eta-infty-B}) for the Bernoulli-chain model at full speed.

\subsection{The error probability of exonuclease proofreading}

Remarkably, it is possible to obtain an expression for the error probability across the crossover from low to high dNTP concentrations.  Instead of approximating the coefficient $b$ by Eq.~(\ref{b-estim}), we go back to its definition~(\ref{b-M}).  Provided that $W^{\rm x}_{+{\rm i}\vert{\rm c}}\ll W^{\rm p}_{+{\rm i}\vert{\rm c}}$ and $W^{\rm p}_{-{\rm i}\vert{\rm c}}\ll W^{\rm x}_{-{\rm i}\vert{\rm c}}$, we get
\be
b \simeq \frac{3\, W^{\rm p}_{+{\rm i}\vert{\rm c}}}{W^{\rm x}_{-{\rm i}\vert{\rm c}} + v_{\rm i}} \, . \label{b-M-estim}
\ee
Now, the partial velocity for incorrect base pair incorporation can be approximated as $v_{\rm i}\simeq c\, v$ using Eq.~(\ref{c-M}) with $v\simeq v_{\rm c}$ so that
\be
v_{\rm i} \simeq \frac{k^{\rm p}_{+{\rm c}\vert{\rm i}}}{K_{{\rm c}\vert{\rm i}}} \, [{\rm dNTP}] \, ,
\label{v_i-estim}
\ee
Replacing into Eq.~(\ref{b-M-estim}) and supposing that the dNTP concentration is still low enough that $Q_{\rm c}\simeq Q_{\rm i}\simeq 1$, the coefficient is evaluated as
\be
b \simeq \frac{3 \, k^{\rm p}_{+{\rm i}\vert{\rm c}}\, K_{{\rm c}\vert{\rm i}} \, [{\rm dNTP}]}{K_{{\rm i}\vert{\rm c}}\left(k^{\rm x}_{-{\rm i}}\, K_{{\rm c}\vert{\rm i}} + k^{\rm p}_{+{\rm c}\vert{\rm i}}\, [{\rm dNTP}]\right)} \, . \label{b-estim2}
\ee
With the coefficient~(\ref{c-estim}), the error probability is again approximated by $\eta \simeq (xy)^{-1} \simeq bc$, whereupon we obtain:
\be
\eta_{\rm M} \simeq \frac{\eta_{\infty,{\rm M}} \, [{\rm dNTP}]}{[{\rm dNTP}] + K^{\rm x}_{\rm M}} 
\label{eta-MM-M}
\ee
with the full speed error probability~(\ref{eta-infty-M}) and the constant
\be
K^{\rm x}_{\rm M} \equiv \frac{k^{\rm x}_{-{\rm i}}\, K_{{\rm c}\vert{\rm i}}}{k^{\rm p}_{+{\rm c}\vert{\rm i}}} \, .
\label{K-x-M}
\ee
Equation~(\ref{eta-MM-M}) with the constants (\ref{eta-infty-M}) and~(\ref{K-x-M}) constitutes the main result of this paper.  In the low speed regime where $[{\rm dNTP}]\ll K^{\rm x}_{\rm M}$, we recover the error probability given by Eq.~(\ref{eta-low-M}).  For $[{\rm dNTP}]\gg K^{\rm x}_{\rm B}$, the error probability~(\ref{eta-infty-M}) at full speed is recovered.  Therefore, Eq.~(\ref{eta-MM-M}) describes the behavior of the error probability in the crossover.  For exonuclease-deficient DNA polymerases, the exonuclease rate constant vanishes $k^{\rm x}_{-{\rm i}}=0$, so that $K^{\rm x}_{\rm M}=0$ and the error probability keeps its maximal value $\eta=\eta_{\infty,{\rm M}}$.  Equation~(\ref{eta-MM-M}) shows that the behavior of the error probability has a Michaelis-Menten reminiscence.

The key point is that the the error probability is able to reach much lower values under the assumptions of the Markov-chain model than under those of the Bernoulli one.  Indeed, for the Bernoulli-chain model, the error probability given by Eqs.~(\ref{eta-MM-M})-(\ref{K-x-M}) would read
\be
\eta_{\rm B} \simeq \frac{\eta_{\infty,{\rm B}} \, [{\rm dNTP}]}{[{\rm dNTP}] + K^{\rm x}_{\rm B}} 
\label{eta-MM-B}
\ee
with the full speed error probability~(\ref{eta-infty-B}) and
\be
K^{\rm x}_{\rm B} \equiv \frac{k^{\rm x}_{-{\rm i}}\, K_{\rm c}}{k^{\rm p}_{+{\rm c}}} \, .
\label{K-x-B}
\ee

Since the polymerization rate slows down after the incorporation of an incorrect base pair $k^{\rm p}_{+{\rm c}\vert{\rm i}} \ll k^{\rm p}_{+{\rm c}\vert{\rm c}}$ and the Michaelis-Menten dissociation constant becomes larger $K_{{\rm c}\vert{\rm i}}>K_{{\rm c}\vert{\rm c}}$~\cite{DPJ91,J93,JJ01b}, the constant~(\ref{K-x-M}) is significantly larger under the assumptions of the Markov-chain model than if the polymerase was insensitive to the previously incorporated nucleotide as in the Bernoulli-chain model, $K^{\rm x}_{\rm M} \gg K^{\rm x}_{\rm B}$.  However, the full speed error probability remains comparable under both types of assumptions, $\eta_{\infty,{\rm M}}\simeq \eta_{\infty,{\rm B}}$. Thanks to the dependence of the polymerization rates on the previously incorporated nucleotide as described by the Markov-chain model, the error probability is thus able to reach much lower values than otherwise.  This proofreading mechanism can be most significant as will be illustrated for the DNA polymerases of T7 viruses and human mitochondria in the following sections.

\section{T7 DNA polymerase}
\label{T7-Pol}

\subsection{Phenomenology}

The kinetics of the exonuclease activity for the wild-type T7 DNA polymerase has been experimentally investigated~\cite{DPJ91,J93}.  The parameter values of the exonuclease activity inferred from the measured data and used for  the present numerical simulations are given in Table~\ref{tab.T7-simul}.  The parameters of the polymerase activity are the same as in the companion paper~\cite{paperI}.  Since there is no complete set of data for every possible pairing, it is the Markov-chain model that is numerically simulated for the T7 DNA polymerase, as in Ref.~\cite{paperI}. The values $[{\rm dNMP}]=10^{-5}$~M and $[{\rm P}]=10^{-4}$~M are used respectively for the concentrations of deoxynucleoside monophosphate and pyrophosphate, which correspond to physiological conditions \cite{T94,H01}.

\begin{table}
\caption{\label{tab.T7-simul} Exo$^+$ T7 DNA polymerase at $20^{\circ}$C: The rate constants of the exonuclease activity used for the numerical simulations and the Markov-chain model.  The rate constants are from Refs.~\cite{DPJ91,J93}.  The other parameters are from the numerical simulations.}
\vspace{5mm}
\begin{center}
\begin{tabular}{|ccc|}
\hline
parameter & value & units \\
\hline
$k^{\rm x}_{-{\rm c}}$ & $0.2$ & s$^{-1}$\\
$k^{\rm x}_{-{\rm i}}$ & $2.3$ & s$^{-1}$\\
$[{\rm dNTP}]_0$ & $2.33\times 10^{-8}$ & M\\
$r^{\rm p}_0=r^{\rm x}_0$ & $0.2$ & nt/s\\
$\eta_{\infty}$ & $1.0\times 10^{-6}$ & nt$^{-1}$\\
$D_{\infty}$ & $1.6\times 10^{-5}$ & nt$^{-1}$\\
$v_{\infty}$ & $288$ & nt/s\\
\hline
\end{tabular}
\end{center}
\end{table} 

\subsection{Numerical and theoretical results}

The kinetics is numerically simulated by using Gillespie's algorithm~\cite{paperI,G76,G77}.  The concentrations of the four nucleotides are supposed to be equal according to Eqs.~(\ref{dNTP}) and~(\ref{dNMP}).  The template is taken as a Bernoulli chain of equal probabilities $\nu_1(n)=\frac{1}{4}$ for $n\in\{{\rm A},{\rm C},{\rm G},{\rm T}\}$.  For every value of dNTP concentration, the growth of $5\times10^3$ chains each of length $10^6$ is numerically simulated and the different quantities of interest are computed by statistical averaging over this sample.   In the following figures, the dots show the results of the numerical simulations, the solid lines those of the Markov-chain model of Section~\ref{M-chain}, and the dashed lines those of the Bernoulli-chain model of Section~\ref{B-chain}.

\begin{figure}[h]
\centerline{\scalebox{0.5}{\includegraphics{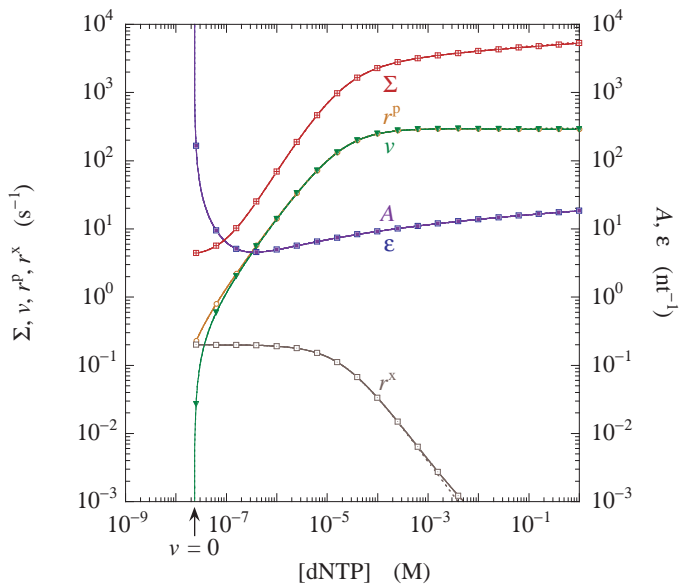}}}
\caption{Exo$^+$ T7 DNA polymerase: Entropy production $\Sigma$ (crossed squares), mean polymerase rate $r^{\rm p}$ (open circles), mean growth velocity $v$ (filled triangles), affinity $A$ (filled squares), free-energy driving force $\epsilon$ (open squares), and mean exonuclease rate $r^{\rm x}$ (dotted squares) versus nucleotide concentration.  The dots are the results of numerical simulations, the solid lines of the Markov-chain model, and the dashed lines of the Bernoulli-chain model.}
\label{fig3}
\end{figure}

\begin{figure}[h]
\centerline{\scalebox{0.5}{\includegraphics{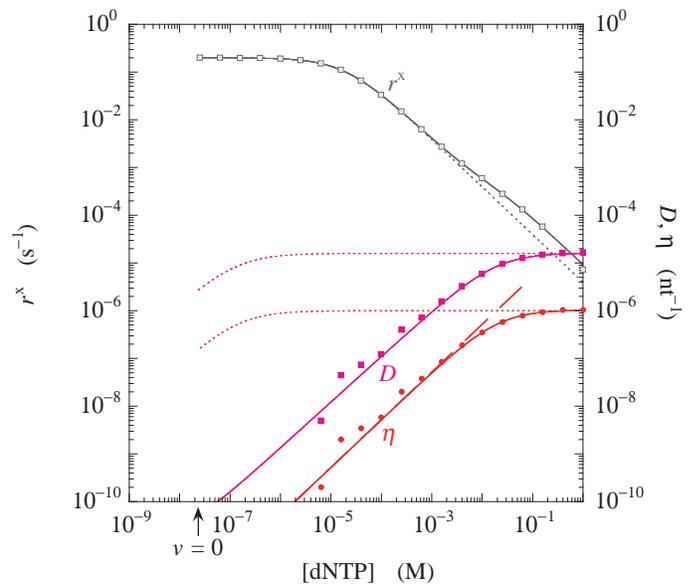}}}
\caption{Exo$^+$ T7 DNA polymerase: Mean exonuclease rate $r^{\rm x}$ (dotted squares), conditional Shannon disorder per nucleotide $D$ (filled squares), and error probability $\eta$ (filled circles) versus nucleotide concentration.  The dots are the results of numerical simulations, the solid lines of the Markov-chain model, and the dashed lines of the Bernoulli-chain model.  The long-dashed line is the behavior of Eq.~(\ref{eta-low-M}).}
\label{fig4}
\end{figure}

In Fig.~\ref{fig3}, we see that the growth velocity vanishes at the critical dNTP concentration given in Table~\ref{tab.T7-simul}, which is very well approximated by both Eqs.~(\ref{dNTP-0-B}) and~(\ref{dNTP-0-M}): $[{\rm dNTP}]_{0,{\rm B}}=[{\rm dNTP}]_{0,{\rm M}} \simeq 2.33\times 10^{-8}$~M.  According to Eq.~(\ref{v-p-x}), the polymerase and exonuclease rates become both equal to the exonuclease rate constant by Eq.~(\ref{rates_p+x_0_B}), as confirmed by the corresponding values in Table~\ref{tab.T7-simul}.  Since the exonuclease activity goes on in spite of the vanishing of the growth velocity, the thermodynamic entropy production does not vanish and takes the positive value $\frac{d_{\rm i}S}{dt}\big\vert_0 \simeq 4.4~R$~s$^{-1}$, estimated by both Eqs.~(\ref{entr-0-B}) and~(\ref{entr-0-M}).  Consequently, the affinity -- which is the entropy production per incorporated nucleotide - and the free-energy driving force per nucleotide are both diverging to infinity if the velocity is vanishing.  In the full speed regime, the exonuclease rate decreases to very small values.  Therefore, the behavior of the exo$^-$ polymerase is recovered.  The growth velocity saturates at its maximal value, which becomes equal to the polymerization rate, while the entropy production increases logarithmically with [dNTP], as the affinity and the free-energy driving force.

In Fig.~\ref{fig4}, the decrease of the exonuclease rate $r^{\rm x}$ is seen to manifest a small shoulder, which is not the case for the Bernoulli-chain model.  The reason is that, as [dNTP] increases, the rate decreases as $r^{\rm x}\simeq 9.8\times 10^{-6}\, [{\rm dNTP}]^{-1}$ by Eq.~(\ref{r^x-infty-M}) of the Markov-chain model, while the Bernoulli-chain one would predict a faster decrease as $r^{\rm x}\simeq 4.0 \times 10^{-6}\, [{\rm dNTP}]^{-1}$ according to Eq.~(\ref{r^x-infty-B}).  

The most prominent result of Fig.~\ref{fig4} is that the error probability and the conditional Shannon disorder per nucleotide take drastically lower values in the Markov-chain model (dots and solid lines) than the Bernoulli one (dashed lines).  If the error probability takes comparable values $\eta_{\infty,{\rm M}}\simeq \eta_{\infty,{\rm B}}\simeq 10^{-6}$ at full speed in both the Bernoulli- and Markov-chain models as expected from Eqs.~(\ref{eta-infty-B}) and~(\ref{eta-infty-M}), in contrast, the error probability becomes much smaller in the Markov-chain model, although it keeps its full speed value down to very small dNTP concentrations in the Bernoulli-chain model.  The behavior observed for the numerical simulations and the Markov-chain model is well described by Eq.~(\ref{eta-low-M}) giving $\eta\simeq 5.18\times 10^{-5}\, [{\rm dNTP}]$, which is the long-dashed line depicted in Fig.~\ref{fig4}.  This low speed behavior and the crossover to the full speed regime are well described by Eqs.(\ref{eta-MM-M})-(\ref{K-x-M}) of the Markov-chain model.  Since the polymerase is slowed down after an incorrect pairing $k^{\rm p}_{+{\rm c}\vert{\rm i}}=0.01\, {\rm s}^{-1} \ll k^{\rm p}_{+{\rm c}\vert{\rm c}}=k^{\rm p}_{+{\rm c}}=300\, {\rm s}^{-1}$, the concentration~(\ref{K-x-M}) where the crossover happens is much larger in the Markov-chain model than the Bernoulli one where it is given by Eq.~(\ref{K-x-B}): $K^{\rm x}_{\rm M}\simeq 1.9\times 10^{-2}\, {\rm M}\gg K^{\rm x}_{\rm B}\simeq 1.5\times 10^{-7}\, {\rm M}$.  Therefore, the error probability keeps its full speed value to a much lower [dNTP] concentration before mildly decreasing in the Bernoulli-chain model (dashed line in Fig.~\ref{fig4}).  For physiological dNTP concentrations~(\ref{physio-dNTP}), the error probability is thus two decades smaller thanks to the dependence of the kinetics on the previously incorporated nucleotide, which cannot be described by Bernoulli-chain models.

\section{Human mitochondrial DNA polymerase}
\label{Hum-Pol}

\subsection{Phenomenology}

For the wild-type human mitochondrial polymerase, the kinetics of the exonuclease activity has been experimentally investigated and reported in Ref.~\cite{JJ01b}.  Table~\ref{tab.Hum-simul} gives the values here used for numerical simulations.  The parameters of the polymerase activity are the same as in the companion paper~\cite{paperI}.  The same values as in the previous Section~\ref{T7-Pol} are taken for the concentrations of deoxynucleoside monophosphate and pyrophosphate.  The parameters of the polymerase activity for the Bernoulli- and Markov-chain models are given in the companion paper~\cite{paperI}.

\begin{table}
\caption{\label{tab.Hum-simul} Exo$^+$ human mitochondrial DNA polymerase $\gamma$ at $37^{\circ}$C: The rate constants of the exonuclease activity used for the numerical simulations.  The rate constants are from Ref.~\cite{JJ01b}.  The other parameters are from the numerical simulations.}
\vspace{5mm}
\begin{center}
\begin{tabular}{|ccc|}
\hline
parameter & value & units \\
\hline
$k^{\rm x}_{-{\rm c}}$ & $0.05$ & s$^{-1}$\\
$k^{\rm x}_{-{\rm i}}$ & $0.4$ & s$^{-1}$\\
$[{\rm dNTP}]_0$ & $1.4\times 10^{-9}$ & M\\
$r^{\rm p}_0=r^{\rm x}_0$ & $0.05$ & nt/s\\
$\eta_{\infty}$ & $1.68\times 10^{-4}$ & nt$^{-1}$\\
$D_{\infty}$ & $1.8\times 10^{-3}$ & nt$^{-1}$\\
$v_{\infty}$ & $34$ & nt/s\\
\hline
\end{tabular}
\end{center}
\end{table} 

\subsection{Numerical and theoretical results}

Gillespie's algorithm is again used to simulate numerically the stochastic process~\cite{paperI,G76,G77}.  The concentrations of the four nucleotides are equal according to Eqs.~(\ref{dNTP}) and~(\ref{dNMP}), while the template is a Bernoulli chain of equal probabilities.  The growth is numerically simulated for $5\times10^3$ chains each of length $10^6$ in order to perform the statistics.  In the following figures,  the results of the numerical simulations are depicted by dots, those of the Markov-chain model of Section~\ref{M-chain} by solid lines, and those of the Bernoulli-chain model of Section~\ref{B-chain} by dashed lines.

\begin{figure}[h]
\centerline{\scalebox{0.5}{\includegraphics{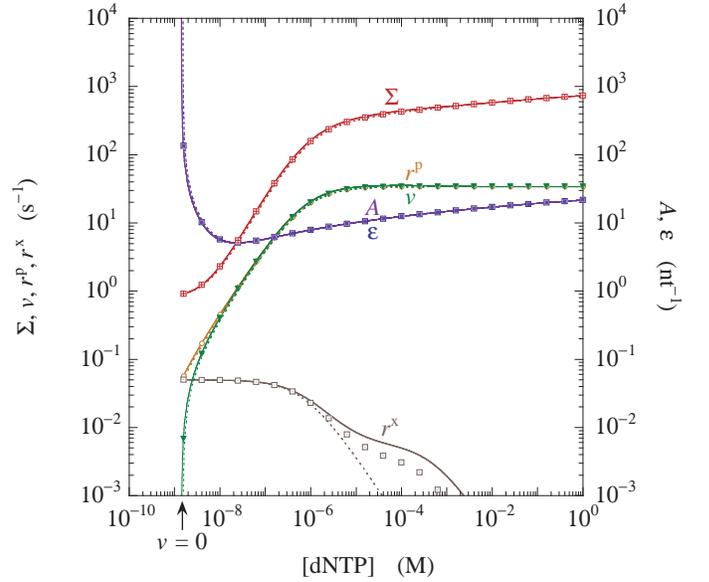}}}
\caption{Exo$^+$ human mitochondrial DNA polymerase: Entropy production $\Sigma$ (crossed squares), mean polymerase rate $r^{\rm p}$ (open circles), mean growth velocity $v$ (filled triangles), affinity $A$ (filled squares), free-energy driving force $\epsilon$ (open squares), and mean exonuclease rate $r^{\rm x}$ (dotted squares) versus nucleotide concentration.  The dots are the results of numerical simulations, the solid lines of the Markov-chain model, and the dashed lines of the Bernoulli-chain model.}
\label{fig5}
\end{figure}

\begin{figure}[h]
\centerline{\scalebox{0.5}{\includegraphics{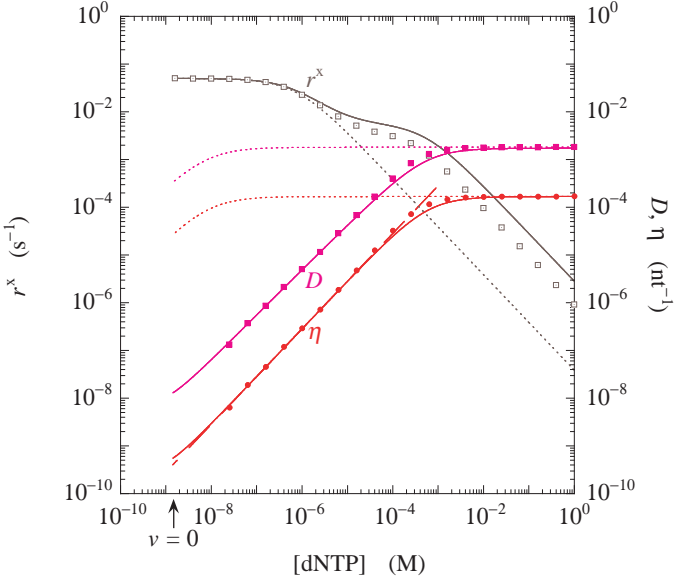}}}
\caption{Exo$^+$ human mitochondrial DNA polymerase: Mean exonuclease rate $r^{\rm x}$ (dotted squares), conditional Shannon disorder per nucleotide $D$ (filled squares), and error probability $\eta$ (filled circles) versus nucleotide concentration.  The dots are the results of numerical simulations, the solid lines of the Markov-chain model, and the dashed lines of the Bernoulli-chain model.  The long-dashed line is the behavior of Eq.~(\ref{eta-low-M}).}
\label{fig6}
\end{figure}

Figure~\ref{fig5} shows the entropy production, the polymerization rate, the growth velocity, the affinity, the free-energy driving force, and the exonuclease rate as a function of dNTP concentration.  The growth velocity is vanishing at the critical dNTP concentration given in Table~\ref{tab.Hum-simul}, which is very well approximated by Eq.~(\ref{dNTP-0-M}) giving $[{\rm dNTP}]_{0,{\rm M}} \simeq 1.42\times 10^{-9}$~M, while Eq.~(\ref{dNTP-0-B}) gives the close value $[{\rm dNTP}]_{0,{\rm B}} \simeq 1.53\times 10^{-9}$~M.   At this critical concentration, the polymerase and exonuclease rates become equal to the value given in Table~\ref{tab.Hum-simul}, which is equal to the exonuclease rate constant expected from Eq.~(\ref{rates_p+x_0_B}).  Accordingly, the thermodynamic entropy remains positive at the value $\frac{d_{\rm i}S}{dt}\big\vert_0 \simeq 0.9~R$~s$^{-1}$ estimated by both Eqs.~(\ref{entr-0-B}) and~(\ref{entr-0-M}).  Hence, the affinity and the free-energy driving force per nucleotide both diverge to infinity if the velocity goes to zero.  As for the T7~DNA~polymerase, the exonuclease rate decreases to very small values in the full speed regime where exonuclease-deficient behavior is recovered.  In this regime, the growth velocity becomes equal to the polymerization rate, reaching the maximal value $v_{\infty}\simeq r^{\rm p}_{\infty}\simeq 34$~nt/s, which is smaller than for the T7~DNA~polymerase, while  the entropy production, the affinity, and the free-energy driving force increase logarithmically with the dNTP concentration.

Figure~\ref{fig6} shows the exonuclease rate, the conditional Shannon disorder, and the error probability corresponding to the quantities in Fig.~\ref{fig5}.  For this exo$^-$ polymerase, the Markov- and Bernoulli-chain models are simplifications of the full kinetics simulated by Gillespie's algorithm, which explains that the solid and dashed lines deviate from the dots for the exonuclease rate $r^{\rm x}$ at large values of dNTP concentration in Fig.~\ref{fig6}.  As [dNTP] increases, the rate decreases as $r^{\rm x}\simeq 3.1\times 10^{-6}\, [{\rm dNTP}]^{-1}$ according to Eq.~(\ref{r^x-infty-M}) of the Markov-chain model, while a faster decrease as $r^{\rm x}\simeq 3.9 \times 10^{-8}\, [{\rm dNTP}]^{-1}$ is given by Eq.~(\ref{r^x-infty-B}) of the Bernoulli-chain model.  Since the exonuclease activity decreases, the growth velocity becomes equal to the polymerase rate as large values of dNTP concentration.

Most remarkably, the error probability and the conditional Shannon disorder per nucleotide are much reduced in the Markov-chain model (dots and solid lines) with respect to the Bernoulli one (dashed lines), as for the exo$^+$~T7~DNA~polymerase.  At full speed, the error probability saturates at its maximal value $\eta_{\infty}\simeq 1.68\times 10^{-4}$ approximated by Eqs.~(\ref{eta-infty-B}) and~(\ref{eta-infty-M}).  However, the error probability is significantly smaller in the Markov- than the Bernoulli-chain model at low dNTP concentration. Indeed, Eq.~(\ref{eta-low-M}) of the Markov-chain model predicts that $\eta\simeq 0.28 \, [{\rm dNTP}]$ giving the long-dashed line depicted in Fig.~\ref{fig6} in agreement with the simulations.  This observed behavior of the error probability is well described by Eqs.~(\ref{eta-MM-M})-(\ref{K-x-M}) of the Markov-chain model.   The Bernoulli-chain model fails to generate this reduction of the error probability because it is not sensitive to the previously incorporated nucleotide.  The Markov-chain model is able to take into account the slowing down of the polymerase after a mismatch thanks to the distinction between the rate constants $k^{\rm p}_{+{\rm c}\vert{\rm i}}=0.3\, {\rm s}^{-1} \ll k^{\rm p}_{+{\rm c}\vert{\rm c}}=37.3\, {\rm s}^{-1}\simeq k^{\rm p}_{+{\rm c}}=34.8\, {\rm s}^{-1}$, which is not possible in the Bernoulli-chain model.  For the same reason, the crossover to the regime with a much lower error probability happens in the Markov-chain model at a larger dNTP concentration than in the Bernoulli because $K^{\rm x}_{\rm M}\simeq 5.4\times 10^{-4}\, {\rm M}\gg K^{\rm x}_{\rm B}\simeq 9.1\times 10^{-9}\, {\rm M}$.  As seen in Fig.~\ref{fig6}, the error probability indeed keeps its full speed value to the much lower concentration $[{\rm dNTP}]\simeq K^{\rm x}_{\rm B}$ in the Bernoulli-chain model (dashed line in Fig.~\ref{fig4}) than in the Markov-chain one where the drop in the error probability already happens for concentrations below $[{\rm dNTP}]\simeq K^{\rm x}_{\rm M}$.  For physiological dNTP concentrations~(\ref{physio-dNTP}), the error probability is thus again two decades smaller thanks to the dependence of the kinetics on the previously incorporated nucleotide, which is the feature of the Markov-chain model.

\section{Conclusion}
\label{Conclusion}

In the present paper, the mechanism of exonuclease proofreading is analyzed in detail using experimental observations from biochemistry \cite{PWJ91,WPJ91,DPJ91,J93,TJ06,JJ01a,JJ01b,LNKC01,LJ06,LK82,EG91,KB00,SUKOOBWFWMG08,RBT08,ZBNS09,DJPW10,BBT12} and theoretical methods already applied to exonuclease-deficient DNA polymerases in the companion paper~\cite{paperI}.

An essential aspect of exonuclease proofreading is the sensitivity of the enzymatic kinetics to mismatches in the base pairing of the previously incorporated nucleotide.  Such mismatches induce a slowing down of the polymerase activity, allowing the DNA strand to jump to the exonuclease domain of the enzyme where the misincorporated nucleotide is cleaved out \cite{J93}.  Such a mechanism would not be possible if the enzyme was memoryless and its rates only depended on the currently incorporated nucleotide, in which case the copy growing on a Bernoullian template would be itself a Bernoulli chain.  If the rates also depend on the previously incorporated nucleotide, the copy forms a Markov chain even if the template is Bernoullian.  A comparison has thus been systematically carried out between the Bernoulli- and Markov-chain models.  For exo$^-$ DNA polymerases, both types of models behave similarly (except close to equilibrium).  In contrast, the difference between both models is drastic in the presence of the exonuclease activity.  If the error probability keeps constant down to low values of dNTP concentration in the Bernoulli-chain model, it decreases significantly in the Markov-chain model, showing how important is the enzymatic memory of previous mismatches to perform exonuclease proofreading.

\begin{figure}[h]
\centerline{\scalebox{0.55}{\includegraphics{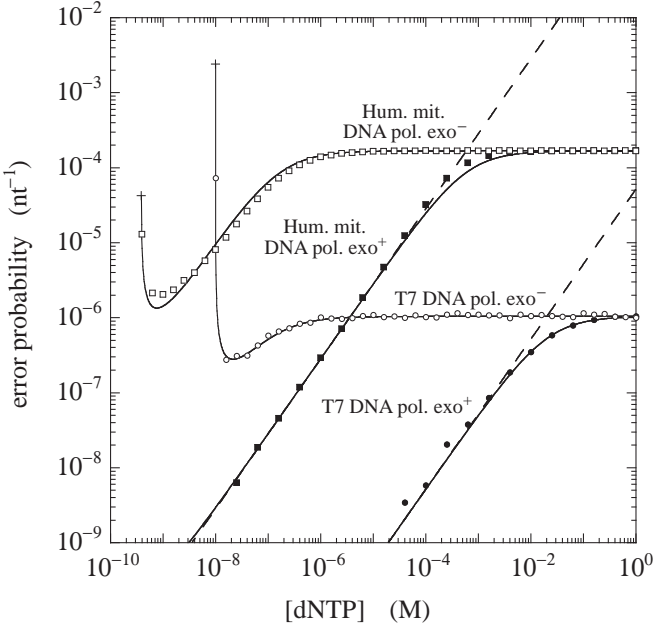}}}
\caption{Error probability versus dNTP concentration for the exo$^-$ (open symbols) and exo$^+$ (filled symbols) DNA polymerases of T7 viruses and human mitochondria.  The dots depict the results of numerical simulations and the lines those of the Markov-chain model.  The long-dashed lines show the behavior of Eq.~(\ref{eta-low-M}). For the exo$^-$ polymerases, the pluses depict the equilibrium values of the error probability.}
\label{fig7}
\end{figure}

In Fig.~\ref{fig7}, the error probability is plotted as a function of dNTP concentration for the different exo$^-$ and exo$^+$ DNA polymerases studied in the companion and present papers~\cite{paperI}.  The results of numerical simulations are depicted as dots and those of the Markov-chain model as solid lines.  The analysis confirms that the replication fidelity is lower for the human mitochondrial DNA polymerase than for the T7 DNA polymerase.  Thanks to the dependence of the rates on the previously incorporated nucleotide taken into account in the Markov-chain model, a large amount of proofreading is achieved by the exonuclease activity.  For physiological dNTP concentrations, we see in Fig.~\ref{fig7} that the error probability undergoes a hundred-fold lowering with respect to the value provided by the kinetic amplification \cite{H74,N75} of the lone polymerase activity at high dNTP concentration in agreement with experimental observations~\cite{WPJ91,DPJ91,J93,JJ01b,LNKC01}.  If the error probability is of the order of $\eta\simeq 10^{-6}$ for an efficient polymerase activity, it will reach values as small as $\eta\simeq 10^{-8}$ with exonuclease proofreading.  This hundred-fold reduction of the error probability is also consistent with the drop in mutation rate from RNA viruses having polymerases devoid of exonuclease activity, to DNA viruses equipped with exonuclease activity~\cite{GEFS09}.  The dependence of the error probability on dNTP concentration is explained thanks to Eq.~(\ref{eta-MM-M}), which describes the crossover from its maximal value if $[{\rm dNTP}]\gg K^{\rm x}_{\rm M}$, down to lower values if $[{\rm dNTP}]\ll K^{\rm x}_{\rm M}$.  In this range of dNTP concentrations, which contains the physiological conditions, the error probability is proportional to the dNTP concentration and it decreases with the concentration.  The crossover concentration $K^{\rm x}_{\rm M}$ given by Eq.~(\ref{K-x-M}) takes a large value precisely thanks to the fact that the polymerization rate is slowed down after the incorporation of an incorrect base pair by $k^{\rm p}_{+{\rm c}\vert{\rm i}}\ll k^{\rm p}_{+{\rm c}\vert{\rm c}}$, while the corresponding Michaelis-Menten dissociation is enhanced because $K_{{\rm c}\vert{\rm i}}\gg K_{{\rm c}\vert{\rm c}}$.  Accordingly, the dependence of the kinetic constants on both the current and previous pairings in the Markov-chain model allows the error probability to decrease with the dNTP concentration.  Consequently, exonuclease proofreading has the advantage over the kinetic amplification of the polymerase activity that the replication fidelity tends to increase if the pool of nucleotides decreases.  The thermodynamic cost is that the enzyme remains away from equilibrium even if the growth velocity vanishes due to the increase of dNMP cleavage.  The dependence~(\ref{eta-MM-M}) of the error probability on nucleotide concentration is a prediction of theory and could be tested experimentally.  

The present theory and methods can be applied as well to other DNA polymerases in order to understand in detail how fidelity depends on important control parameters such as the substrate concentrations.  Furthermore, the analytical results should allow us to calculate the error probability of exonuclease proofreading using modern computational approaches.  Indeed, the rate and dissociation constants in Eqs.~(\ref{eta-infty-M}) or~(\ref{K-x-M}) can be determined by using Arrhenius' law of kinetics in terms of the free-energy landscape of the enzyme-DNA complex along its reaction pathway and conformational changes thanks to the computational methods \cite{FGW05,XGBWW08,ROW10}.

An open issue is that an error probability of about $10^{-8}$ for exonuclease proofreading, as evaluated in the present paper for efficient DNA polymerases under physiological conditions, would limit the genome size to $10^8$ nucleotides according to the theory of quasispecies by Eigen and Schuster \cite{ES77,ES78a,ES78b}.  The fact is that other proofreading mechanisms such as the postreplication DNA mismatch repair \cite{L74,SR83,IPBM06} are in action to further reduce the error probability in higher eukaryotes having genome sizes as large as $10^{10}$ nucleotides \cite{GEFS09}.

\vskip 0.5 cm

\begin{acknowledgments}
The author is grateful to D. Andrieux, D. Bensimon, J. England, D. Lacoste, Y. Rondelez, and S.~A.~Rice for helpful discussions, remarks, and support during the elaboration of this work.
This research is financially supported by the Universit\'e Libre de Bruxelles, the FNRS-F.R.S., and the Belgian Federal Government under the Interuniversity Attraction Pole project P7/18 ``DYGEST".
\end{acknowledgments}

\begin{widetext}

\appendix

\section{Equations for kinetics and thermodynamics}
\label{AppA}

For the reaction network depicted in Fig.~\ref{fig1}, the kinetic equations ruling the time evolution of the probabilities that the growing copy has respectively the sequences $m_1\cdots m_l$ and $m_1\cdots m_l m_{\rm l+1}{\rm P}$ are given by
\bea
&&\frac{d}{dt}\, {\cal P}_t\left(m_1 \cdots m_l \qquad\quad\ \atop n_1\, \cdots \, n_l \, n_{l+1}\cdots\right) = \ k^{\rm p}_{+m_l m_{l-1}\atop \ \, n_l \, n_{l-1}}
\, {\cal P}_t\left(m_1 \cdots m_l{\rm P} \qquad\ \ \atop n_1\, \cdots \, n_l \, n_{l+1}\cdots\right)
+\sum_{m_{l+1}} k_{-m_{l+1} m_l\atop \ \, n_{l+1} \, n_l} 
\, {\cal P}_t\left(m_1 \cdots m_l m_{l+1}{\rm P} \qquad\ \ \atop n_1\, \cdots \, n_l \, n_{l+1}\, n_{l+2}\cdots\right)
\nonumber\\
&&\qquad\qquad\qquad\qquad\qquad\qquad +\
k^{\rm x}_{+m_l m_{l-1}\atop \ \, n_l \, n_{l-1}}[m_l]
\, {\cal P}_t\left(m_1 \cdots m_{l-1} \qquad\ \, \atop n_1\, \cdots \, n_{l-1} \, n_l\cdots\right)
+\sum_{m_{l+1}} k^{\rm x}_{-m_{l+1} m_l\atop \ \, n_{l+1} \, n_l} 
\, {\cal P}_t\left(m_1 \cdots m_l m_{l+1} \qquad\quad \ \atop n_1\, \cdots \, n_l \, n_{l+1}\, n_{l+2}\cdots\right)
\nonumber\\
&&\qquad - \left( k^{\rm p}_{-m_l m_{l-1}\atop \ \, n_l \, n_{l-1}}[{\rm P}]
+\sum_{m_{l+1}} k_{+m_{l+1} m_l\atop \ \, n_{l+1} \, n_l}[m_{l+1}{\rm P}]
+ k^{\rm x}_{-m_l m_{l-1}\atop \ \, n_l \, n_{l-1}}
+\sum_{m_{l+1}} k^{\rm x}_{+m_{l+1} m_l\atop \ \, n_{l+1} \, n_l}[m_{l+1}]\right)  
{\cal P}_t\left(m_1 \cdots m_l \qquad\quad \ \atop n_1\, \cdots \, n_l \, n_{l+1}\cdots\right) 
\label{kin_eq_1}
\eea
and
\bea
\frac{d}{dt}\, {\cal P}_t\left(m_1 \cdots m_l m_{l+1}{\rm P} \qquad\ \ \atop n_1\, \cdots \, n_l \, n_{l+1}\, n_{l+2}\cdots\right) &=& 
k_{+m_{l+1} m_l\atop \ \, n_{l+1} \, n_l}[m_{l+1}{\rm P}]
\, {\cal P}_t\left(m_1 \cdots m_l \qquad\quad\ \atop n_1\, \cdots \, n_l \, n_{l+1}\cdots\right)
+k^{\rm p}_{-m_{l+1} m_l\atop \ \, n_{l+1} \, n_l}[{\rm P}]
\, {\cal P}_t\left(m_1 \cdots m_l m_{l+1} \qquad\quad\ \atop n_1\, \cdots \, n_l \, n_{l+1}\, n_{l+2}\cdots\right)
\nonumber\\
&-& \left( k_{-m_{l+1} m_l\atop \ \, n_{l+1} \, n_l} + k^{\rm p}_{+m_{l+1} m_l\atop \ \, n_{l+1} \, n_l}\right)  
{\cal P}_t\left(m_1 \cdots m_l m_{l+1}{\rm P} \qquad\ \ \atop n_1\, \cdots \, n_l \, n_{l+1}\, n_{l+2}\cdots\right)
\label{kin_eq_2}
\eea
in terms of the rates~(\ref{ntb_rate})-(\ref{nmd_rate}) for $l=1,2,3,...$.  In Eq.~(\ref{kin_eq_1}) for the probability of a copy ending with a monophosphate group, the gain terms are due to polymerization, dNTP dissociation, dNMP binding, and dNMP dissociation, while the loss terms are due to depolymerization, dNTP binding, dNMP dissociation, and dNMP binding.  In Eq.~(\ref{kin_eq_2}) for the probability of a copy ending with a triphosphate group, the gain terms are due to dNTP binding and depolymerization, and the loss terms to dNTP dissociation and polymerization.  For $l=1$ in Eq.~(\ref{kin_eq_1}) and $l=0$ in Eq.~(\ref{kin_eq_2}), the symbols $m_0$ and $n_0$ stand for the empty set: $m_0=n_0=\emptyset$.  For $l=0$, Eq.~(\ref{kin_eq_1}) should be replaced by
\bea
\frac{d}{dt}\, {\cal P}_t\left(\emptyset\qquad\quad \ \atop n_1\, n_2 \, \cdots\right) &=&
\sum_{m_1} k_{-m_1 \emptyset \atop \ \ n_1\, \emptyset} 
\, {\cal P}_t\left(m_1{\rm P} \qquad \ \atop n_1\, n_2 \, \cdots\right)
+\sum_{m_1} k^{\rm x}_{-m_1 \emptyset \atop \ \ n_1\, \emptyset} 
\, {\cal P}_t\left(m_1 \qquad\quad \atop n_1\, n_2 \, \cdots\right) \nonumber\\
&-&\left( \sum_{m_1} k_{+m_1\emptyset\atop \ \ n_1\, \emptyset}[m_1{\rm P}]+ \sum_{m_1} k^{\rm x}_{+m_1\emptyset\atop \ \ n_1\, \emptyset}[m_1]\right)
{\cal P}_t\left(\emptyset\qquad\quad \ \atop n_1\, n_2\cdots\right) \, .
\label{kin_eq_0}
\eea
The equations (\ref{kin_eq_1})-(\ref{kin_eq_0}) preserve the total probability.
We notice that the sequence probabilities are proportional to the corresponding concentrations in a dilute solution.

Under the assumption~(\ref{MM-hyp}), the kinetic equations (\ref{kin_eq_1})-(\ref{kin_eq_2}) reduce to the following kinetic equation
\bea
\frac{d}{dt}\, P_t\left(m_1 \cdots m_l \qquad\quad\ \atop n_1\, \cdots \, n_l \, n_{l+1}\cdots\right) &=& 
W_{+m_l m_{l-1}\atop \ \, n_l \, n_{l-1}} 
\, P_t\left(m_1 \cdots m_{l-1} \qquad\ \ \atop n_1\, \cdots \, n_{l-1} \, n_l\cdots\right)
+\sum_{m_{l+1}} W_{\quad\, -m_{l+1} m_l\atop n_{l+2}\, n_{l+1} \, n_l}
\, P_t\left(m_1 \cdots m_l m_{l+1} \qquad\ \ \atop n_1\, \cdots \, n_l \, n_{l+1}\, n_{l+2}\cdots\right)
\nonumber\\
&-& \left( W_{\quad\, -m_l m_{l-1}\atop n_{l+1}\, n_l \, n_{l-1}}
+\sum_{m_{l+1}} W_{+m_{l+1} m_l\atop \ \, n_{l+1}\, n_l}\right)  
P_t\left(m_1 \cdots m_l \qquad\quad \ \atop n_1\, \cdots \, n_l \, n_{l+1}\cdots\right) \, ,
\label{kin_eq}
\eea
ruling the sum of probabilities
\be
P_t\left(m_1 \cdots m_l \qquad\quad\ \atop n_1\, \cdots \, n_l \, n_{l+1}\cdots\right)
\equiv
{\cal P}_t\left(m_1 \cdots m_l \qquad\quad\ \atop n_1\, \cdots \, n_l \, n_{l+1}\cdots\right) 
+\sum_{m_{l+1}}
 {\cal P}_t\left(m_1 \cdots m_l m_{l+1}{\rm P} \qquad\ \ \atop n_1\, \cdots \, n_l \, n_{l+1}\, n_{l+2}\cdots\right) \, .
\label{prob_sum}
\ee
The rates of Eq.~(\ref{kin_eq}) are given by Eqs.~(\ref{W+})-(\ref{W-}).  The total probability is also preserved by the kinetic equations~(\ref{kin_eq}).

The rates appearing in Eqs.~(\ref{v-p-x}) and~(\ref{r_dNTP}) can be expressed in terms of the rates and probabilities ruled of the kinetic equation~(\ref{kin_eq}) as
\be
r^{\rho} = \nu^{\rho}\, \sum_l\sum_{m_1\cdots m_l}\left[W^{\rho}_{\quad\, +m_l m_{l-1}\atop n_{l+1}\, n_l \, n_{l-1}} \, P_t\left(m_1 \cdots m_{l-1} \qquad\  \atop n_1\, \cdots \, n_{l-1} \, n_l\cdots\right)- W^{\rho}_{\quad\, -m_l m_{l-1}\atop n_{l+1}\, n_l \, n_{l-1}}\, P_t\left(m_1 \cdots m_l \qquad\quad\ \atop n_1\, \cdots \, n_l \, n_{l+1}\cdots\right)\right] \, 
\label{rates_p+x}
\ee
with the stoichiometric coefficient $\nu^{\rm p}=+1$ for the polymerase activity $\rho=$~p, and the stoichiometric coefficient $\nu^{\rm x}=-1$ for the exonuclease one $\rho=$~x.

Now, the thermodynamic entropy production is given by \cite{P55,S76,N79,LVN84,JQQ04,G04}
\bea
\frac{1}{R}\frac{d_{\rm i}S}{dt} &=&\sum_{\rho={\rm p},{\rm x}} \sum_l \sum_{m_1\cdots m_l}
\left[W^{\rho}_{+m_l m_{l-1}\atop \ \, n_l \, n_{l-1}} 
\, P_t\left(m_1 \cdots m_{l-1} \qquad\ \ \atop n_1\, \cdots \, n_{l-1} \, n_l\cdots\right)
-W^{\rho}_{\quad\, -m_l m_{l-1}\atop n_{l+1}\, n_l \, n_{l-1}}
\, P_t\left(m_1 \cdots m_l \qquad\quad\ \ \atop n_1\, \cdots \, n_l \, n_{l+1}\cdots\right)\right]
\nonumber\\
&&\qquad\qquad\qquad\qquad\qquad\qquad
\times \ln\frac{W^{\rho}_{+m_l m_{l-1}\atop \ \, n_l \, n_{l-1}} 
\, P_t\left(m_1 \cdots m_{l-1} \quad\ \ \ \atop n_1\, \cdots \, n_{l-1} \, n_l\cdots\right)}
{W^{\rho}_{\quad\, -m_l m_{l-1}\atop n_{l+1}\, n_l \, n_{l-1}}
\, P_t\left(m_1 \cdots m_l \qquad\ \ \atop n_1\, \cdots \, n_l \, n_{l+1}\cdots\right)} \geq 0 \, ,
\label{entr-A}
\eea
which includes the contributions of polymerase and exonuclease activities.

\section{Thermodynamics of the Bernoulli-chain model}
\label{AppB}

For the Bernoulli-chain model defined in terms of the rates~(\ref{Wx+_B})-(\ref{Wx-_B}) besides the polymerization-depolymerization rates, the thermodynamic entropy production~(\ref{entr-A}) becomes
\be
\frac{1}{R}\frac{d_{\rm i}S}{dt} =\sum_{\rho={\rm p},{\rm x}} 
\left\{\left[W^{\rho}_{+{\rm c}} - W^{\rho}_{-{\rm c}} (1-\eta)\right] \, \ln\frac{W^{\rho}_{+{\rm c}}}{W^{\rho}_{-{\rm c}} (1-\eta)}
+ \left(3\, W^{\rho}_{+{\rm i}} - W^{\rho}_{-{\rm i}}\, \eta\right) \, \ln\frac{3\, W^{\rho}_{+{\rm i}}}{W^{\rho}_{-{\rm i}} \, \eta} \right\} \geq 0 \, .
\label{entr-B}
\ee
Separating the terms with the error probability in the logarithm and using Eq.~(\ref{B-velocity}) for the mean growth velocity $v$, we obtain the expression~(\ref{entr-prod}) for the thermodynamic entropy production in terms of the free-energy driving force
\be
\epsilon = \frac{1}{v} \sum_{\rho={\rm p},{\rm x}} 
\left\{\left[W^{\rho}_{+{\rm c}} - W^{\rho}_{-{\rm c}} (1-\eta)\right] \, \ln\frac{W^{\rho}_{+{\rm c}}}{W^{\rho}_{-{\rm c}}}
+ \left(3\, W^{\rho}_{+{\rm i}} - W^{\rho}_{-{\rm i}}\, \eta\right) \, \ln\frac{W^{\rho}_{+{\rm i}}}{W^{\rho}_{-{\rm i}}} \right\} \geq 0 \, ,
\label{eps-B}
\ee
and the conditional Shannon disorder per nucleotide 
\be
D(\omega\vert\alpha)= -(1-\eta)\, \ln(1-\eta) -\eta \, \ln\frac{\eta}{3} \, .
\label{D-B}
\ee

\section{Thermodynamics of the Markov-chain model}
\label{AppC}

For the Markov-chain model, the thermodynamic entropy production~(\ref{entr-A}) is given by
\bea
\frac{1}{R}\frac{d_{\rm i}S}{dt} =\sum_{\rho={\rm p},{\rm x}} && \Big\{
\left[W^{\rho}_{+{\rm c}\vert{\rm c}}\, \mu({\rm c}) - W^{\rho}_{-{\rm c}\vert{\rm c}} \, \mu({\rm c}\vert{\rm c}) \, \mu({\rm c})\right] \, \ln\frac{W^{\rho}_{+{\rm c}\vert{\rm c}}}{W^{\rho}_{-{\rm c}\vert{\rm c}} \, \mu({\rm c}\vert{\rm c})} \nonumber\\
&& +3
\left[W^{\rho}_{+{\rm c}\vert{\rm i}}\, \mu({\rm i}) - W^{\rho}_{-{\rm c}\vert{\rm i}} \, \mu({\rm i}\vert{\rm c}) \, \mu({\rm c})\right] \, \ln\frac{W^{\rho}_{+{\rm c}\vert{\rm i}}\, \mu({\rm i})}{W^{\rho}_{-{\rm c}\vert{\rm i}} \, \mu({\rm i}\vert{\rm c}) \, \mu({\rm c})} \nonumber\\
&& +3
\left[W^{\rho}_{+{\rm i}\vert{\rm c}}\, \mu({\rm c}) - W^{\rho}_{-{\rm i}\vert{\rm c}} \, \mu({\rm c}\vert{\rm i}) \, \mu({\rm i})\right] \, \ln\frac{W^{\rho}_{+{\rm i}\vert{\rm c}}\, \mu({\rm c})}{W^{\rho}_{-{\rm i}\vert{\rm c}} \, \mu({\rm c}\vert{\rm i}) \, \mu({\rm i})} \nonumber\\
&& +9
\left[W^{\rho}_{+{\rm i}\vert{\rm i}}\, \mu({\rm i}) - W^{\rho}_{-{\rm i}\vert{\rm i}} \, \mu({\rm i}\vert{\rm i}) \, \mu({\rm i})\right] \, \ln\frac{W^{\rho}_{+{\rm i}\vert{\rm i}}}{W^{\rho}_{-{\rm i}\vert{\rm i}} \, \mu({\rm i}\vert{\rm i})}\geq 0 \, ,
\label{entr-M}
\eea
in terms of the conditional probabilities $\mu(p'\vert p)$ and tip probabilities $\mu(p)$ of the Markov chain with $p,p'={\rm c}$ or~${\rm i}$.

\vskip 4 cm

.
\end{widetext}


\end{document}